\documentclass[journal]{article}

\usepackage{tikz}
\usepackage{xifthen}

\pgfdeclarelayer{bg}    
\pgfsetlayers{bg,main}  
\definecolor{beige}{RGB}{245, 245, 220}
\newcommand*\circled[1]{\tikz[baseline=(char.base)]{
            \node[shape=circle,draw,inner sep=1pt] (char) {#1};}}

\usetikzlibrary{calc, arrows, fit, positioning, patterns, decorations.pathreplacing, shapes}
\tikzstyle{dash} = [dashed, -latex,>=latex]
\tikzstyle{line} = [draw, -latex,>=latex]
\tikzstyle{smallbox} = [draw, minimum size=5.0mm, font=\scriptsize]
\tikzstyle{box} = [draw, minimum size=10.0mm]
\tikzstyle{roundbox} = [draw, circle, inner sep=0pt, minimum size=3mm]
\tikzstyle{clamped} = [draw, fill=black, minimum size=0.15cm]
\tikzstyle{msgcircle} = [shape=circle, draw, inner sep=0pt, minimum size=3.5mm, fill=white, font=\scriptsize]

\newcommand{\msg}[6]{
      \ifthenelse{\isin{#1}{left} \AND \isin{#2}{down}}{
            \coordinate (anchor) at ($({#3})!{#5}!({#4})$);
            \node[msgcircle, xshift=-5.0mm] at (anchor) {#6};
            \node[xshift=-1.5mm] at (anchor) {$\downarrow$};
      }{}
      \ifthenelse{\isin{#1}{right} \AND \isin{#2}{down}}{
            \coordinate (anchor) at ($({#3})!{#5}!({#4})$);
            \node[msgcircle, xshift=5.0mm] at (anchor) {#6};
            \node[xshift=1.5mm] at (anchor) {$\downarrow$};
      }{}

      \ifthenelse{\isin{#1}{down} \AND \isin{#2}{right}}{
            \coordinate (anchor) at ($({#3})!{#5}!({#4})$);
            \node[msgcircle, yshift=-5.5mm] at (anchor) {#6};
            \node[yshift=-2.0mm] at (anchor) {$\rightarrow$};
      }{}
      \ifthenelse{\isin{#1}{up} \AND \isin{#2}{right}}{
            \coordinate (anchor) at ($({#3})!{#5}!({#4})$);
            \node[msgcircle, yshift=5.5mm] at (anchor) {#6};
            \node[yshift=2.0mm] at (anchor) {$\rightarrow$};
      }{}

      \ifthenelse{\isin{#1}{down} \AND \isin{#2}{left}}{
            \coordinate (anchor) at ($({#3})!{#5}!({#4})$);
            \node[msgcircle, yshift=-5.5mm] at (anchor) {#6};
            \node[yshift=-2.0mm] at (anchor) {$\leftarrow$};
      }{}
      \ifthenelse{\isin{#1}{up} \AND \isin{#2}{left}}{
            \coordinate (anchor) at ($({#3})!{#5}!({#4})$);
            \node[msgcircle, yshift=5.5mm] at (anchor) {#6};
            \node[yshift=2.0mm] at (anchor) {$\leftarrow$};
      }{}

      \ifthenelse{\isin{#1}{left} \AND \isin{#2}{up}}{
            \coordinate (anchor) at ($({#3})!{#5}!({#4})$);
            \node[msgcircle, xshift=-5.0mm] at (anchor) {#6};
            \node[xshift=-1.5mm] at (anchor) {$\uparrow$};
      }{}
      \ifthenelse{\isin{#1}{right} \AND \isin{#2}{up}}{
            \coordinate (anchor) at ($({#3})!{#5}!({#4})$);
            \node[msgcircle, xshift=5.0mm] at (anchor) {#6};
            \node[xshift=1.5mm] at (anchor) {$\uparrow$};
      }{}
}

\usepackage{amsfonts,amsmath,amssymb}

\renewcommand{\d}[1]{\operatorname{d}\!{#1}}
\renewcommand{\exp}[1]{\operatorname{exp}\!\left({#1}\right)}
\renewcommand{\log}[1]{\operatorname{log}\!\left({#1}\right)}

\newcommand{\E}[1]{\mathbb{E}\!\left[{#1}\right]}

\usepackage{todonotes}
\usepackage[normalem]{ulem}
\usepackage{cancel}
\usepackage{soul}
\usepackage{hyperref}
\usepackage[all]{hypcap}
\hypersetup{pdfborder={0 0 0}}
\usepackage{enumitem}
\usepackage{pdfpages}
\usepackage{color}
\usepackage{hhline}
\usepackage[linesnumbered,ruled]{algorithm2e}

\renewcommand{\d}[1]{\operatorname{d}\!{#1}}

\SetKwComment{Comment}{$\triangleright$\ }{}

\begin{document}


\title{A Probabilistic Modeling Approach to One-Shot Gesture Recognition}
\author{Anouk~van~Diepen,
        Marco~Cox
        and~Bert~de~Vries,
\thanks{Anouk van Diepen and Marco Cox are with the Department of Electrical Engineering,
Eindhoven University of Technology, PO Box 513, 5600 MB Eindhoven, the
Netherlands (e-mail: a.v.diepen@tue.nl; m.g.h.cox@tue.nl).}
\thanks{Bert de Vries is with GN Hearing BV, Het Eeuwsel 6, 5612 AS Eindhoven, the
Netherlands and with the Department of Electrical Engineering, Eindhoven
University of Technology, PO Box 513, 5600 MB Eindhoven, the Netherlands
(email: bdevries@ieee.org).}
}
\maketitle

\begin{abstract}
Gesture recognition enables a natural extension of the way we currently interact with devices. Commercially available gesture recognition systems are usually pre-trained and offer no option for customization by the user. In order to improve the user experience, it is desirable to allow end users to define their own gestures. This scenario requires learning from just a few training examples if we want to impose only a light training load on the user. To this end, we propose a gesture classifier based on a hierarchical probabilistic modeling approach.
In this framework, high-level features that are shared among different gestures can be extracted from a large labeled data set, yielding a prior distribution for gestures. When learning new types of gestures, the learned shared prior reduces the number of required training examples for individual gestures.
We implemented the proposed gesture classifier for a Myo sensor bracelet and show favorable results for the tested system on a database of 17 different gesture types. Furthermore, we propose and implement two methods to incorporate the gesture classifier in a real-time gesture recognition system.
\end{abstract}

\section{Introduction}\label{sec:intro}
Gesture recognition, i.e., recognition of pre-defined gestures by arm or hand movements, enables a natural extension of the way we currently interact with devices \cite{horsley_wave_2016}. With the increasing amount of human-machine interactions, alternative user interfaces will become more important. Gesture recognition will not replace existing technology, but will solve issues that are not addressed by current input devices: think about a surgeon who wants to display patient information during surgery, but also about less pressing issues, such as scrolling through a recipe while cooking or declining a call without having to pick up the telephone.

Commercially available gesture recognition systems are usually pre-trained: the developers specify a set of gestures, and the user is provided with an algorithm that can recognize just these gestures. To improve the user experience, it is often desirable to allow users to define their own gestures. In that case, the user needs to train the recognition system herself by performing example gestures. Crucially, this scenario requires learning gestures from just a few training examples in order to avoid overburdening the user; this scheme is also known as one-shot learning \cite{lake_one-shot_2014}.

\begin{figure}[t]
   \centering
\scalebox{1}{\includegraphics[width=0.485\textwidth]{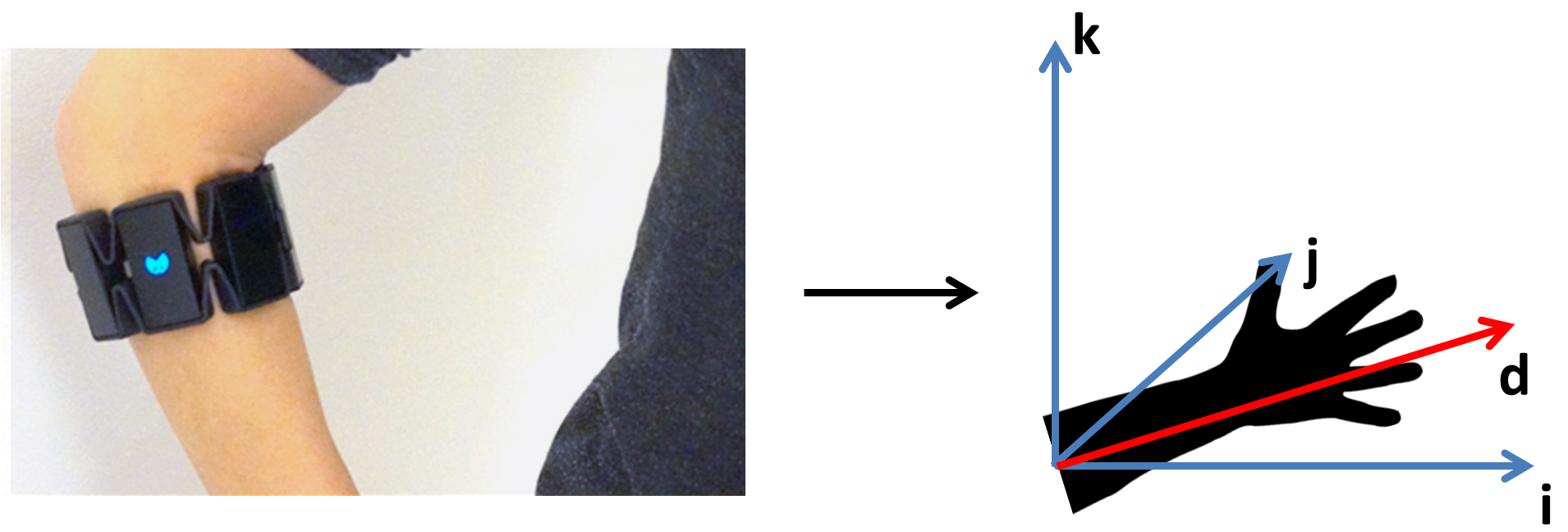}}
\caption{The Myo sensor bracelet used to measure the direction of the arm. The direction of the arm is represented by a three-dimensional unit vector $\mathbf{d}$.}\label{fig:myo}
\end{figure}

The first step towards a one-shot gesture recognition system is to create a one-shot gesture \emph{classifier}. The task of this classifier is to assign gesture class labels to sequences of movement measurements. Training such a classifier usually requires many training examples or even hand-crafted features for each gesture class. One-shot learning seems to come naturally to humans: people are capable of using prior experiences when learning new tasks. This principle can be translated into an algorithm by using a Bayesian modeling approach. The Bayesian approach facilitates learning of prior knowledge, which reduces the number of required examples for learning a new class. Casting both learning and classification as inference tasks in these (probabilistic) models yields a principled way to design and evaluate algorithm candidates.

In this paper we present a new one-shot trainable gesture classifier based on a hierarchical probabilistic modeling approach for directional data of arm movements. The proposed gesture classifier allows users to define their own gesture classes by giving examples of gestures. By employing a Bayesian approach, we can make use of a (large) set of gestures to train a `prior' for gestures in our model. As a result, new gesture classes can be added to the classifier by means of very few class-specific examples. In Section \ref{sec:algorithm} the classifier is explained in more detail.

The gesture classifier is designed to work on directional data of the arm. In principle, the specific measurement mechanics are not relevant for the classifier. To test our classifier, we measure gestures using a Myo sensor bracelet (see Fig. \ref{fig:myo}), which measures the orientation of the arm using inertial and magnetic sensors. Our classifier extracts the direction of the arm from the orientation measurements, and uses this as input to classify new measurements (see Section \ref{sec:measurements}).

To test our classifier, we have collected a database of gesture measurements, consisting of 17 different types of gestures performed by 6 different users. In Section \ref{sec:Results} we evaluate the accuracy on this dataset, compare the classifier with a classifier based on dynamic time warping, and analyze the influence of including prior information.

In practice, providing a user with only a working gesture classifier is not enough. A working gesture recognition system must also be able to detect gestures in a continuous stream of data. This issue often leads to an additional burden on the end user, e.g., when she is required to mark the start and end of a gesture by pushing a button \cite{liu_uwave:_2009}. In Section \ref{sec:on} we propose and implement two real-time methods for incorporating the gesture classifier in a working gesture recognition system. The first approach uses an on-line version of the gesture classifier, the second approach uses a ``key gesture" to trigger the activation of the gesture classifier.

In order to put related work in proper context, we defer our discussion of related literature on gesture recognition to Section~\ref{sec:relwork}. The conventions used for mathematical notation are summarized in Appendix \ref{app:notation}.

\section{Classifier design} \label{sec:algorithm}
The task of a classifier is to decide to which gesture class a new sequence of measurements belongs, given previously labeled gestures. This task can be split into two subtasks: there has to be a method to save information contained in previous gestures (learning) and a method to obtain the most likely class of a new gesture measurement (recognition).
Under the probabilistic modeling approach, both learning and recognition are problems of probabilistic inference in the same generative model. This section starts with an intuitive explanation of the classifier; after this, we will specify the generative model, and define learning and recognition as tasks in this model.

\subsection{Intuitive explanation}
\begin{figure}[t]
   \centering
\scalebox{1}{ \includegraphics[width=0.485\textwidth]{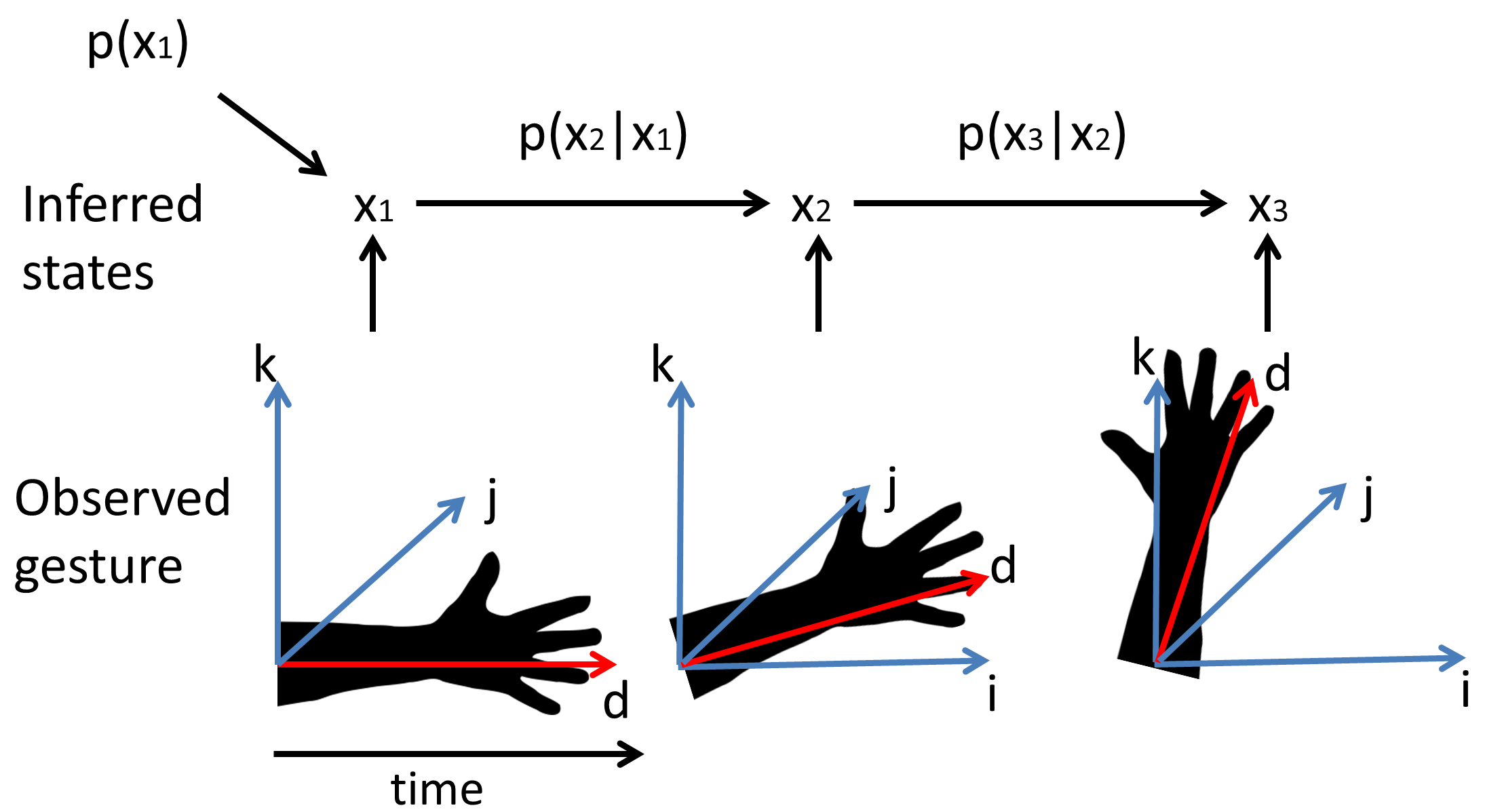}}
\caption{The general idea of modeling gestures using hidden Markov models. The direction of the arm for a certain time step is converted into a discrete state represented by $\mathbf{x}_t$. The transition probabilities between the states specify how likely a movement from one state to another is. By learning and saving this, the characteristics of a gesture are captured.}\label{fig:concept}
\end{figure}

Our proposed gesture classifier is based on Bayesian hidden Markov models. The general idea behind using hidden Markov models is that the measured direction of the arm can be mapped onto discrete states. Instead of saving the discrete states, the algorithm learns the so-called transition probabilities between the states (see Fig. \ref{fig:concept}). A transition probability can be viewed as the probability that the arm will go from one direction to another: in other words, it corresponds to the probability of a movement. These transition probabilities are unique for every distinguishable type (class) of gestures. Therefore, after the algorithm has learned the transition probabilities for every gesture class, it can later (during the recognition task) evaluate the probability that a new measurement was generated by a specific class. This concept has already been successfully applied to gesture classification in the past \cite{mantyla_hand_2000, yamato_recognizing_1992, campbell_invariant_1996, chen_hand_2003}. One of the advantages of hidden Markov models is that they can automatically handle temporal variations, which makes them very suitable for gesture recognition.

\begin{figure}[t]
   \centering
 \scalebox{1}{\includegraphics[width=0.485\textwidth]{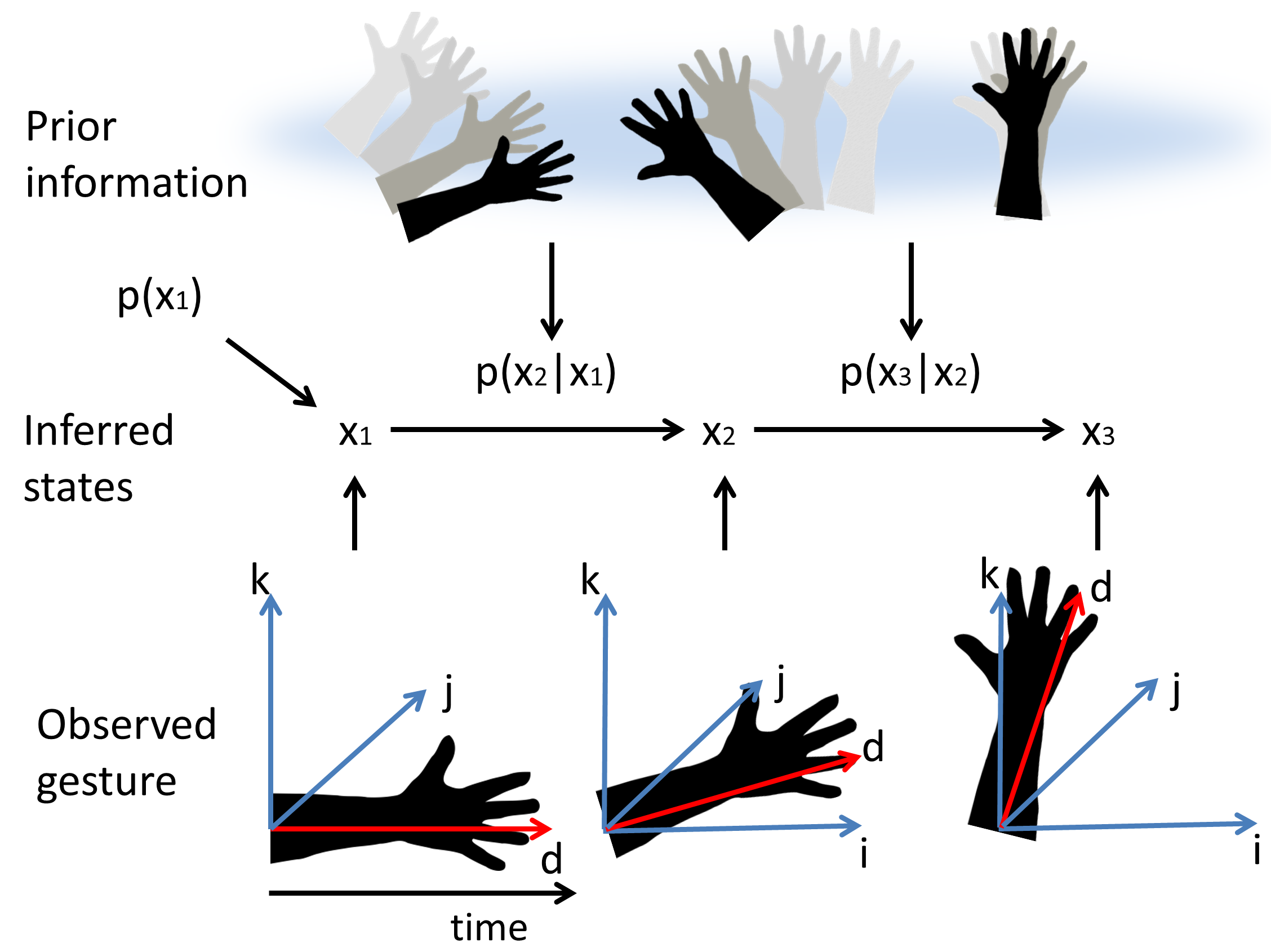}}
\caption{The advantage of using Bayesian hidden Markov models. State transition probabilities are not only learned from measurements, but can also make use of prior information about gestures in the form of a prior distribution.}\label{fig:concept_prior}
\end{figure}

In contrast to earlier work, we use \emph{Bayesian} hidden Markov models as the basis of a \emph{hierarchical} gesture recognition model. This allows us to incorporate prior information about gestures, in the form of a prior probability distribution over the transition probabilities (see Fig. \ref{fig:concept_prior}). The transition probabilities for a specific gesture class can then be learned by combining the prior information and the measurements. The prior distribution, which captures shared properties of the set of all considered gestures, helps to reduce the need for training examples when learning a new gesture class.

\subsection{Generative model specification}
A generative probabilistic model is a joint probability distribution over all (hidden and observed) variables in the system under study.

Let $\mathbf{y} = (y_1,...,y_T)$ with $y_t \in \{1,2,...,N\}$ be a time series of measurements corresponding to a single gesture with underlying characteristics $\theta$. The characteristics are unique for gestures of type (class) $k$. We can capture these dependencies by the probability distribution
\begin{equation}\label{eq:compgenmod}
p(\mathbf{y},\theta, k ) = \underbrace{p(\mathbf{y}|\theta)}_{\substack{\text{dynamical} \\ \text{model}}} \cdot \underbrace{p(\theta|k)}_{\substack{\text{ gesture} \\ \text{characteristics}}} \cdot \underbrace{p(k)}_{\substack{\text{gesture} \\ \text{class index}}}\,.
\end{equation}

Because the measurement sequence is temporally correlated and noisy, we specify $p(\mathbf{y}|\theta)$ as a hidden Markov model (HMM). As a result of this, the dynamic model factorizes as
\begin{equation}
p(\mathbf{y},\mathbf{x}|\theta) = p(x_1|\boldsymbol{\pi})\prod_{t=1}^{T}p(y_t|x_t,\mathbf{C})\prod_{t=1}^{T-1} p(x_{t+1}|x_{t},\mathbf{A}),
\label{eq:genhmm}
\end{equation}
where we have introduced a hidden state sequence $\mathbf{x}=(x_1,...,x_T)$ with $x_t \in \{1,2,...,M\}$  and $\theta \triangleq \{\mathbf{A},\mathbf{C},\boldsymbol{\pi}\}$  with $\mathbf{A} \in \mathbb{R}^{M \times M}$ , $\mathbf{C} \in \mathbb{R}^{N \times M}$ and $\boldsymbol\pi \in \mathbb{R}^{M}$. The elements of $\mathbf{A}$, $\mathbf{C}$ and $\boldsymbol{\pi}$ are defined in the following way:
\begin{align*}
\pi_i &= p(x_{1} = i)\\
A_{ij} &= p(x_{t+1} =i| x_{t}=j )\\
C_{ij} &= p(y_{t} =i| x_{t}=j )\,.
\end{align*}
$A_{ij}$ is called a \emph{transition} probability because it describes the transition between two consecutive states, $C_{ij}$ is the \emph{emission} probability, and $\boldsymbol{\pi}$ is the initial state probability. The conditional probability distributions in Eq.~\ref{eq:genhmm} can then be written as:
\begin{align*}
p(x_1|\boldsymbol\pi)&= \operatorname{Cat}(x_1|\boldsymbol\pi)\\
p(x_{t+1}| x_{t},\mathbf{A})&=\operatorname{Cat}(x_{t+1}|(A_{1x_t},...,A_{Mx_t}))\\
p(y_t| x_{t},\mathbf{C})&=\operatorname{Cat}(y_t|(C_{1x_t},...,C_{Nx_t}))\,,
\end{align*}
where $\operatorname{Cat}$ represents a categorical distribution.
As a result of the model choice, $p(\theta|k)$ factorizes as
\begin{equation}
p(\theta|k) = p(\mathbf{A}|k)p(\mathbf{C}|k)p(\boldsymbol\pi|k)\,.
\end{equation}
The class-conditional distributions over the parameters are assumed to be described by
\begin{align*}
p(\mathbf{A}|k)&=\prod_{j=1}^M \operatorname{Dir}((A_{1j},...,A_{Mj})|\mathbf{h}_j^{(\mathbf{A},k)})\\
p(\mathbf{C}|k)&=\prod_{j=1}^M \operatorname{Dir}((C_{1j},...,C_{Nj})|\mathbf{h}^{(\mathbf{C},k)}_{j})\\
p(\boldsymbol\pi|k) &= \operatorname{Dir}(\boldsymbol\pi|\mathbf{h}^{(\boldsymbol \pi,k)})\,,
\end{align*}
where $\mathbf{h}_j^{(\mathbf{A},k)} \in \mathbb{R}^{M}$, $\mathbf{h}_j^{(\mathbf{C},k)}\in \mathbb{R}^{N}$, $\mathbf{h}^{(\boldsymbol\pi,k)} \in \mathbb{R}^M$ parameterize the Dirichlet distributions. The Dirichlet distribution is a conjugate prior for the categorical distribution. In the literature, $\mathbf{h}_j^{(\cdot,k)}$ are often called hyperparameters since they  parameterize the parameter distributions. Each class $k$ gets its own set of hyperparameters $\mathbf{h}_j^{(\cdot,k)}$. Before training, the hyperparameters are initialized to equal values for all classes.

We define a uniform categorical distribution over the gesture classes $k \in \{1,2,...,K\}$,  which means that each gesture is a priori equally likely:
\begin{align}
p(k) = \operatorname{Cat}\left(k \,\middle\vert\, \left(\frac{1}{K},...,\frac{1}{K}\right)\right)\,.
\end{align}
The class variable $k$ can be seen as a selector that determines from which HMM the gesture is generated. The set of equations in this subsection completely define the generative model.

\subsection{Learning}\label{ssec:learning}
During learning, a method is needed to save the properties of a gesture class $k$. We do this by learning the distribution over the parameters $\theta$ for each gesture class $k$.

Let $\mathbf{y}^{(k)}$ be a measurement from known class $k$. We define a dataset $\mathcal{D}_k = \{\mathbf{y}^{(k)}_1,...,\mathbf{y}^{(k)}_L\}$ containing multiple gestures from a single gesture class $k$. Calculation of the (posterior) distribution over $\theta$ corresponds to substituting observed data for the measurements $\mathcal{D}_k$ and class index $k$ in the generative model (Eq.~\ref{eq:compgenmod}), followed by applying the rules of probability theory to obtain the needed quantity:

\begin{equation}\label{eq:examp_learn}
p(\theta|\mathcal{D}_k,k) = \frac{p(\mathcal{D}_k, \theta, k)}{\int p(\mathcal{D}_k, \theta, k) \d \theta}\,.
\end{equation}

Since we want to train with a very small number of gestures in dataset $\mathcal{D}_k$, this dataset might not be sufficient to obtain a posterior for $\theta$ that gives good results during recognition.
For this reason, we choose to learn the posterior distribution for $\theta$ using a two-step approach.

In the first step, a prior distribution for $\theta$ is constructed. This prior distribution can be obtained in various ways. We have chosen to construct one that captures the common characteristics that are shared among \textit{all} gestures. We define a dataset $\mathcal{D}$ consisting of one measurement from each gesture class: $\mathcal{D} = \{\mathbf{y}^{(1)},...,\mathbf{y}^{(K)}\}$ with $K$ the total number of gesture classes. First, a general prior distribution is learned using dataset $\mathcal{D}$:
\begin{equation}\label{eq:prior}
p(\theta|\mathcal{D}) = \frac{p(\mathcal{D}, \theta)}{\int p(\mathcal{D}, \theta) \d \theta}\,.
\end{equation}
This distribution is independent of $k$.

In the second step, the parameter distribution for $\theta$ is updated for a specific gesture class, using the previously learned $p(\theta|\mathcal{D})$ and the available measurements from a specific gesture class $\mathcal{D}_k = \{\mathbf{y}^{(k)}_1,...,\mathbf{y}^{(k)}_L\}$:
\begin{equation}\label{eq:learning}
p(\theta| \mathcal{D}, \mathcal{D}_k , k) = \frac{p(\mathcal{D}_k|\theta,k)p(\theta|\mathcal{D})p(k)}{\int p(\mathcal{D}_k, \theta, k|\mathcal{D}) \d \theta}\,.
\end{equation}
This distribution $p(\theta| \mathcal{D}, \mathcal{D}_k , k)$ is class dependent.

Alternatively, we could have also chosen to construct a prior distribution ourselves, based on ``common sense" about gestures, e.g., it is very unlikely that an arm movement goes from one state to another, without passing a state in between these states. Learning the prior distribution from gesture data should result in similar constraints.\\

In practice, exact evaluation of Eq.~\ref{eq:prior} and Eq.~\ref{eq:learning} is intractable for our model due to the integral in the denominator. We use variational Bayesian inference to approximate the distributions of Eqs.~\ref{eq:prior} and \ref{eq:learning}. In variational inference, the posterior distribution $p(\mathbf{x},\theta|\mathcal{D},k)$ over hidden states and parameters is approximated by a ``variational'' distribution $q(\mathbf{x},\theta)$ \cite{jordan_introduction_1999}. For computational simplicity, we assume that the variational distribution factorizes by a functional ``mean field'' division as
\begin{equation}\label{eq:mean-field-assumption}
q(\mathbf{x},\theta) = q(\mathbf{x})q(\mathbf{A})q(\mathbf{C})q(\pmb{\pi})\,.
\end{equation}
Note that this assumption does in general not hold for the true posterior distribution $p(\mathbf{x},\theta|\mathcal{D},k)$, so variational inference with assumption Eq.~\ref{eq:mean-field-assumption} leads to an approximation of the true posterior. In order to get a good approximation, a distance measure between this distribution and the true posterior distribution $p(\mathbf{x},\theta|\mathcal{D},k)$ needs to be minimized. The variational (Bayesian) inference method uses the ``exclusive'' Kullback-Leibler (KL) divergence as the distance measure. Minimization of the KL divergence results in a set of update equations that has been derived in \cite{mackay_ensemble_1997} and needs to be updated until convergence.

We then approximate the posterior parameter distribution with the variational parameter distributions:
\begin{equation}
p(\theta|\mathcal{D},k)\approx q(\mathbf{A})q(\mathbf{C})q(\boldsymbol{\pi})\,.
\end{equation}
The approximations for Eq.~\ref{eq:prior} and Eq.~\ref{eq:learning} can both be obtained using this same recipe.

\subsection{Recognition}\label{ssec:rec}
During recognition, the task of the algorithm is to find the class label $k^*$ with the highest probability for an unlabeled measurement $\mathbf{y}$:
\begin{equation}
k^*(\mathbf{y}) = \arg \max_{k} p(k|\mathbf{y})\,.
\end{equation}
Since we assume that each gesture has the same \emph{a priori} probability $p(k)$, it follows that $p(k|\mathbf{y}) \propto p(\mathbf{y}|k)$. The marginal likelihood (the ``evidence'' for class $k$) $p(\mathbf{y}|k)$ can be calculated by marginalization over the model parameters:
\begin{align}
p(\mathbf{y}|k) = \int \underbrace{p(\mathbf{y}|\theta)}_{\text{HMM}}\underbrace{p(\theta|\mathcal{D},\mathcal{D}_k, k)}_{\text{learned distribution}} \d \theta\,.
\label{eq:marg_like}
\end{align}
Again, integration over the complete parameter distribution is intractable and variational inference could be used to approximate the integral. However, in a practical gesture recognition system this step needs to be executed in real-time and therefore needs to be as fast as possible. In order to fast-track the computation, we follow the approach proposed in \cite{beal_variational_2003}, in which the posterior parameter distributions are approximated by their expected values. With this approximation, the marginal likelihood evaluates to
\begin{align}
p(\mathbf{y}|k) &\approx \int  p(\mathbf{y}|\theta)\delta (\theta-\hat\theta_k) \d \theta \notag \\
&= p(\mathbf{y} | \hat\theta_k)\,,
\label{eq:marg_approx}
\end{align}
where $\hat \theta_k = \E{\theta|k} = \int \theta p(\theta|\mathcal{D}, \mathcal{D}_k,k) \d \theta$. In principle, Eq.~\ref{eq:marg_approx} can be evaluated exactly using the forward algorithm \cite{rabiner_tutorial_1989}, which has a time complexity of $\mathcal{O}(M^2T)$. We are also interested in automating inference tasks such as Eq.~\ref{eq:marg_approx}.
In the next section we present a message passing approach that automates inference of Eq.~\ref{eq:marg_approx}.

\subsection{Message passing in Forney-style factor graphs}\label{ssec:FFG}
In this work we have chosen to implement the generative model, learning, and recognition using Forney-style factor graphs (FFGs), which provide an efficient framework to solve inference problems \cite{forney_codes_2001}. FFGs also provide a visually intuitive overview of both the generative model and related inference algorithms. This section provides a brief introduction to Forney-style factor graphs.
An FFG represents a joint probability distribution or generative model by an undirected graph. The factors of the joint probability distribution are represented by nodes, and edges represent the (random) variables. An edge is connected to a node if the factor has the corresponding random variable in its argument list.

\begin{figure}[t]
\centering \scalebox{1}{

\begin{tikzpicture}
        [node distance=20mm,auto,>=stealth']
        \begin{scope}
           \node[] (etc_begin){$\dots$};
	\node[box,right of =etc_begin ](first_state){};
	\node[smallbox,right of =first_state ](eq_second){$=$};
	\node[right of = eq_second] (etc_end){$\dots$};
	
	\node[box,below of = eq_second,node distance=25mm](emiss_sec){};	
	
	\node[clamped, below of = emiss_sec](o_2){};
	
	\node[below of = first_state, node distance = 10mm](){$p(x_{t}|x_{t-1},\mathbf{\hat A}_k)\hspace{5mm}$};

	\node[right of = emiss_sec, node distance = 17mm](){$p(y_{t}|x_{t},\mathbf{\hat C}_k)$};
	\node[left of = emiss_sec, node distance = 15 mm](cmodel){};

	\node[above of = eq_second, node distance = 10mm](niks){};
	\node[smallbox, left of = niks, node distance = 9mm](chain_C){$=$};
	\node[above of = etc_begin, node distance= 10mm](etc_C){$\dots$};
	\node[above of = etc_end, node distance = 10mm](etc_C_2){$\dots$};

	\node[smallbox, above of = first_state](chain_A){$=$};
	\node[left of = chain_A](etc_A){$\dots$};
	\node[above of = etc_end](etc_A_2){$\dots$};

	\path[line] (etc_begin) edge[-] node[anchor=south]{$x_{t-1}$} (first_state); 
	\path[line] (first_state) edge[-] node[anchor=south]{} (eq_second);               
	\path[line] (eq_second) edge[-] node[anchor=south]{$x_t$} (etc_end);       

	\path[line] (eq_second)edge[-]node[anchor=west]{} (emiss_sec);       

	\path[line](emiss_sec)edge[-]node[anchor= east]{$y_{t}$}(o_2);      

	\draw[dashed] (chain_C) node[anchor=south]{} -- ($(emiss_sec)-(0.9,0)$) -> (emiss_sec);        
	\draw[dashed] (chain_C) node[anchor=south]{} -- ($(emiss_sec)-(0.9,0)$) -> (emiss_sec);    
	\path[draw] (chain_C) edge[dashed] node[anchor=south]{} (etc_C);     
	\path[draw] (chain_C) edge[dashed] node[anchor=south]{$\mathbf{\hat C}_k$} (etc_C_2);                                          
	
	\path[draw] (chain_A) edge[dashed] node[anchor=south]{} (first_state); 
	\path[draw] (chain_A) edge[dashed] node[anchor=south]{} (etc_A);     
	\path[draw] (chain_A) edge[dashed] node[anchor=south]{$\mathbf{\hat A}_k$} (etc_A_2);  

        \end{scope}
    \end{tikzpicture}
}
\caption{A (Forney-style) factor graph representation of the joint probability distribution in Eq.~\ref{eq:genone}. A small black node indicates a variable that is observed. In this framework, we need equality nodes (indicated by ``=") as branching points in case a variable appears in more than two factors.}\label{fig:gen_mod}
\end{figure}
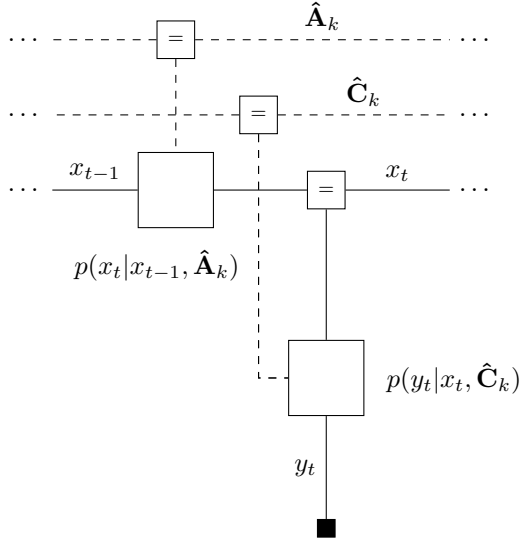

Consider Eq.~\ref{eq:genhmm} for one time step $t$,
\begin{equation}\label{eq:genone}
p(y_t,x_t|x_{t-1},\mathbf{\hat A}_k,\mathbf{\hat C}_k) = p(y_t|x_t, \mathbf{\hat C}_k)p(x_t|x_{t-1},\mathbf{\hat A}_k)\,,
\end{equation}
where $\mathbf{\hat A}_k$ and $\mathbf{\hat C}_k$ represent $\mathbb{E}[\mathbf{A}|k]$ and $\mathbb{E}[\mathbf{C}|k]$ respectively as defined in Section \ref{ssec:rec}, and $y_t$ is observed. A factor graph representation of Eq.~\ref{eq:genone} is drawn in Fig. \ref{fig:gen_mod}. Edges of observed variables are connected to black nodes and equality nodes (indicated by ``=") represent computationally proper branching points for variables that appear in more than two factors \cite{loeliger_introduction_2004, loeliger_factor_2007}.

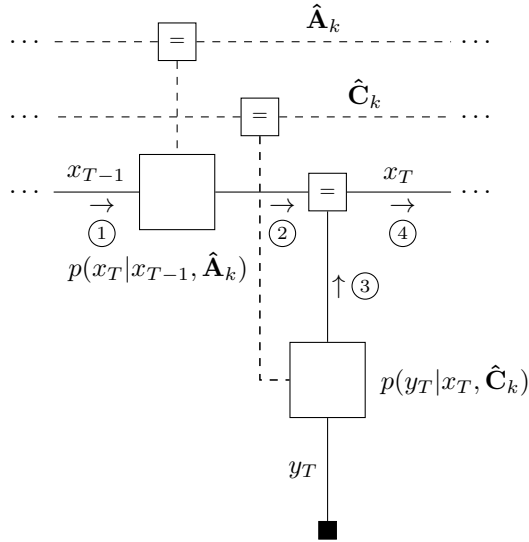
\begin{figure}[t]
   \centering
    \scalebox{1}{

\begin{tikzpicture}
        [node distance=20mm,auto,>=stealth']
        \begin{scope}
           \node[] (etc_begin){$\dots$};
	\node[box,right of =etc_begin ](first_state){};
	\node[smallbox,right of =first_state ](eq_second){$=$};
	\node[right of = eq_second] (etc_end){$\dots$};
	
	\node[box,below of = eq_second,node distance=25mm](emiss_sec){};	
	
	\node[clamped, below of = emiss_sec](o_2){};
	
	\node[below of = first_state, node distance = 10mm](){$p(x_{T}|x_{T-1},\mathbf{\hat A}_k)\hspace{5mm}$};

	\node[right of = emiss_sec, node distance = 17mm](){$p(y_{T}|x_{T},\mathbf{\hat C}_k)$};
	\node[left of = emiss_sec, node distance = 15 mm](cmodel){};

	\node[above of = eq_second, node distance = 10mm](niks){};
	\node[smallbox, left of = niks, node distance = 9mm](chain_C){$=$};
	\node[above of = etc_begin, node distance= 10mm](etc_C){$\dots$};
	\node[above of = etc_end, node distance = 10mm](etc_C_2){$\dots$};

	\node[smallbox, above of = first_state](chain_A){$=$};
	\node[left of = chain_A](etc_A){$\dots$};
	\node[above of = etc_end](etc_A_2){$\dots$};

	\path[line] (etc_begin) edge[-] node[anchor=south]{$x_{T-1}$} (first_state); 
			\msg{down}{right}{etc_begin}{first_state}{0.5}{$1$}
	\path[line] (first_state) edge[-] node[anchor=south]{} (eq_second);
			\msg{down}{right}{first_state}{eq_second}{0.7}{$2$}               
	\path[line] (eq_second) edge[-] node[anchor=south]{$x_T$} (etc_end);       
			\msg{down}{right}{eq_second}{etc_end}{0.5}{$4$}       
	\path[line] (eq_second)edge[-]node[anchor=west]{} (emiss_sec);       
			\msg{right}{up}{eq_second}{emiss_sec}{0.5}{$3$}       
	\path[line](emiss_sec)edge[-]node[anchor= east]{$y_{T}$}(o_2);            

	\draw[dashed] (chain_C) node[anchor=south]{} -- ($(emiss_sec)-(0.9,0)$) -> (emiss_sec);        
	\draw[dashed] (chain_C) node[anchor=south]{} -- ($(emiss_sec)-(0.9,0)$) -> (emiss_sec);    
	\path[draw] (chain_C) edge[dashed] node[anchor=south]{} (etc_C);     
	\path[draw] (chain_C) edge[dashed] node[anchor=south]{$\mathbf{\hat C}_k$} (etc_C_2);                                          
	
	\path[draw] (chain_A) edge[dashed] node[anchor=south]{} (first_state); 
	\path[draw] (chain_A) edge[dashed] node[anchor=south]{} (etc_A);     
	\path[draw] (chain_A) edge[dashed] node[anchor=south]{$\mathbf{\hat A}_k$} (etc_A_2);  

        \end{scope}
    \end{tikzpicture}
}
\caption{Factor graph representation of Eq.~\ref{eq:genone} for the last time step $t=T$, including the messages needed for recognition.}\label{fig:rec_inf}
\end{figure}

In order to execute the classification task, the probability $p(\mathbf{y}|k)$ needs to be evaluated (Eq.~\ref{eq:marg_approx}). This formula can be worked out as
\begin{align*}
p(\mathbf{y}|k) &= \int \sum_{x_T}\sum_{x_1,...,x_{T-1}}p(\mathbf{y},\mathbf{x}|\theta)\delta (\theta-\hat \theta_k)\d \theta\\
&= \sum_{x_T}p(y_T|x_T,\mathbf{\hat C}_k) \sum_{x_{T-1}} p(x_T|x_{T-1},\mathbf{\hat A}_k) \times\\
\end{align*}
\vspace{-1.2cm}
\begin{align*}
    &\quad \underbrace{\sum_{x_1,...,x_{T-2}}p(x_1|\boldsymbol{\hat \pi}_k)\prod_{t=1}^{T-1}p(y_t|x_t,\mathbf{\hat C}_k)\prod_{t=2}^{T-1} p(x_{t}|x_{t-1},\mathbf{\hat A}_k)}_{\circled{1}}\\ \label{eq:inference_example}
 & = \sum_{x_T}\overbrace{\underbrace{p(y_T|x_T,\mathbf{\hat C}_k)}_{\circled{3}} \underbrace{ \sum_{x_{T-1}} p(x_T|x_{T-1},\mathbf{\hat A}_k)\circled{1}}_{\circled{2}}}^{\circled{4}}\,.
\end{align*}
We can see that $\circled{1}$ only depends on terms to the left of the factor graph in Fig. \ref{fig:rec_inf} and is only a function of $x_{T-1}$. $\circled{1}$ can therefore be represented as a message on the edge of $x_{T-1}$, flowing from the left to the right. The remaining computations $\circled{2}$, $\circled{3}$ and $\circled{4}$ can similarly be represented as messages on edges (see Fig. \ref{fig:rec_inf}).
Incoming message $\circled{1}$ is obtained from the outgoing message of the previous time slice (not shown) of the graph. 
If we store message computation rules for common node functions in a lookup table, then Eq.~\ref{eq:marg_approx} can be efficiently executed by making use of these stored rules.

The particular scheme discussed above is called sum-product message passing and can only be used when exact inference is possible. For variational inference, which is needed for learning the parameter distributions, a different message passing procedure called ``variational message passing'' is needed \cite{dauwels_variational_2007}. The specific message update equations that we used in this paper are derived in Appendix \ref{sec:var}.

\section{Experimental evaluation}\label{sec:Results}

To evaluate the performance of our classifier, we measure gestures using the Myo sensor bracelet \cite{noauthor_myo_nodate}. In this section we report on experimental evaluation results.

\subsection{Measuring gestures}\label{sec:measurements}

The bracelet contains a nine-axis inertial measurement unit (IMU) containing a three-axis gyroscope, three-axis accelerometer and three-axis magnetometer.  The software on the bracelet itself filters these measurements and delivers orientation data of the arm, which is a filtered combination of the IMU data. The details of the bracelet's preprocessing filters are undisclosed. This section describes how the direction of the arm is extracted from Myo's orientation data stream.

\begin{figure}[t]
   \centering
 \scalebox{1}{\includegraphics[width=0.35\textwidth]{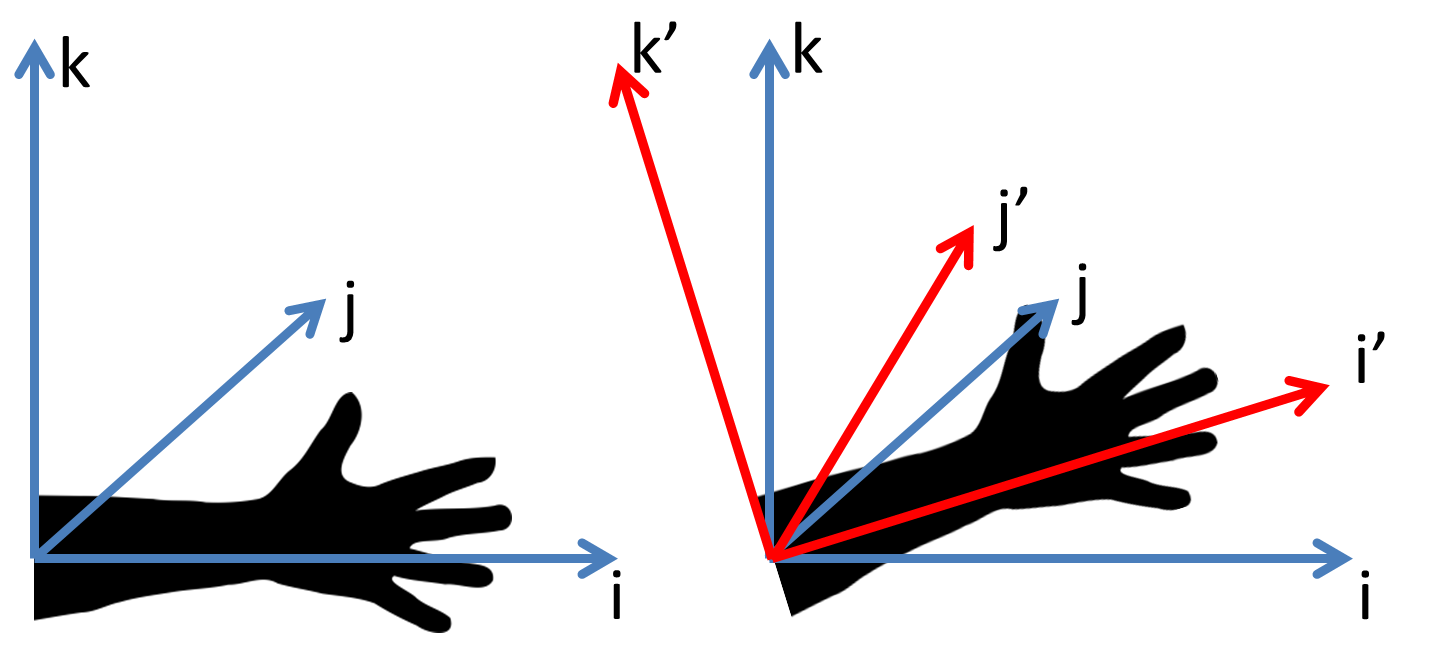}}
\caption{On the left, the the global frame of reference of the arm. On the right, a new orientation of the arm with a rotated reference frame that is rigidly attached to the arm (red). Orientation is the rotation between the global reference frame (blue) and the rotated frame (red).}\label{fig:orientation}
\end{figure}

Orientation is an imaginary rotation around a global reference frame (see Fig. \ref{fig:orientation}). Orientation in three dimensions can be encoded in several ways, for example by rotation matrices, Euler angles or unit quaternions. Each of these formalisms is equivalent and has its own advantages.

The software of the Myo bracelet encodes the orientation of the arm by a unit quaternion for computational reasons. According to Euler's rotation theorem, every rotation of a rigid body in three dimensions can be described by a three-dimensional unit vector $\mathbf{v}$, and an angle $\phi$. A unit quaternion is simply another way to encode this as
\begin{align*}
\mathbf{\dot{q}} &= \left(\cos \left(\frac{\phi}{2}\right), \mathbf{v}\sin \left(\frac{\phi}{2}\right)\right)\,.
\end{align*}

Because unit quaternions only describe the relative orientation of the arm and not the absolute position, in our test application gestures are restricted to be performed with an extended arm. Furthermore, the user always needs to hold his arm in a ``synchronization" position (pointing the arm to the right) before making a gesture.

In order to obtain the current arm direction $\mathbf{d}$, we rotate the arm's synchronization position $\mathbf{d}_{sync} = (1,0,0)$ by the measured compensated quaternion $\mathbf{\dot{q}}_{comp}$, leading to
\begin{align*}
\mathbf{d}= \mathbf{\dot{q}}_{comp}\mathbf{d}_{sync}\mathbf{\dot{q}}_{comp}^{-1}\,,
\end{align*}
where we make use of the fact that any vector $\mathbf{v}$ can be rotated by a quaternion $\mathbf{\dot{q}}$ through application of $\mathbf{\dot{q}} \mathbf{v} \mathbf{\dot{q}}^{-1}$.

Since our HMM assumes measurements are chosen from a discrete alphabet, we need to quantize the raw direction $\mathbf{d}$ to a discrete observation $y \in \{1,2,\ldots,N\}$. This is achieved by a simple vector quantization algorithm. We have set a (manually tuned) grid of $N$ basis vectors $\mathbf{b}_n$ in the space of movements (see top-left in Fig.~\ref{fig:types_gest}) and round the measured direction $\mathbf{d}$ to the nearest basis vector, yielding a new measurement at every time step, given by
$$
y = \arg\min_{n \in N} \| \mathbf{d} - \mathbf{b}_n \| \,.
$$

\subsection{Data set}
\begin{figure}[t]
   \centering
 \scalebox{1}{\includegraphics[width=0.5\textwidth]{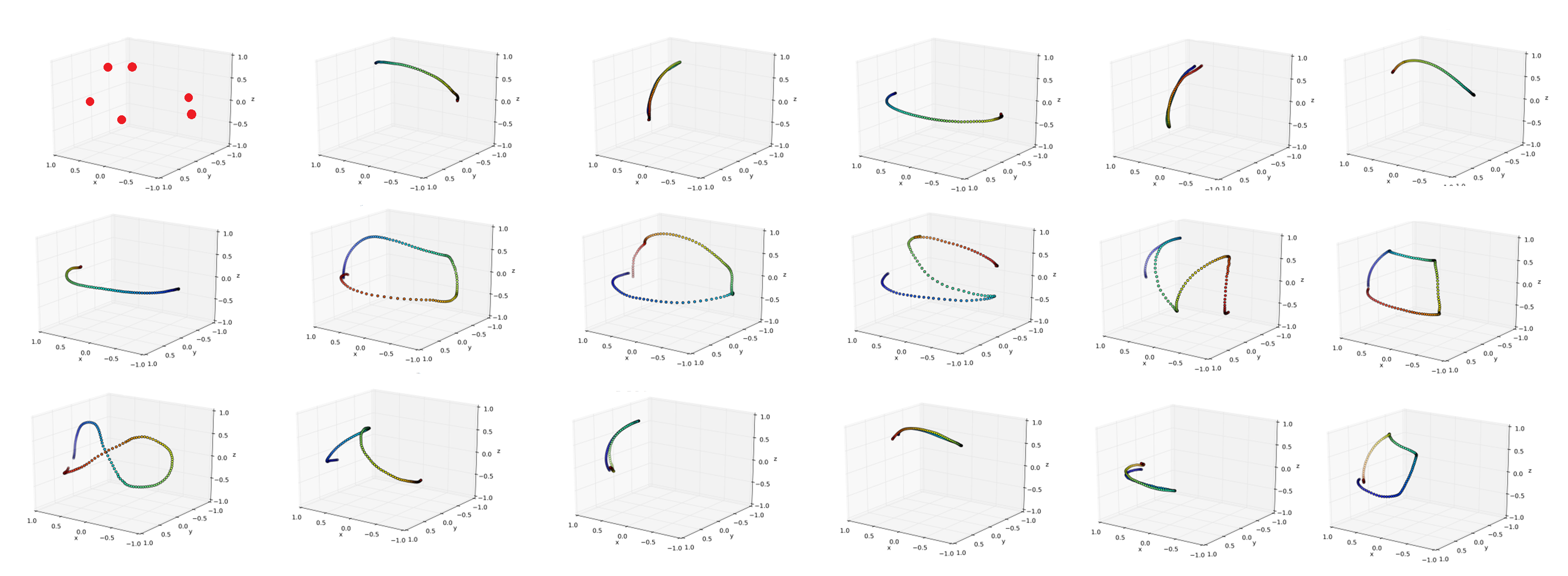}}
\caption{Visualization of the 6 quantization points (top left) and one example of each gesture class. The motion of the gestures is from blue to red. }\label{fig:types_gest}
\end{figure}

We built a gesture database using the Myo sensor bracelet \cite{database}. The orientation signal was sampled at 6.7 Hz, converted into the direction of the arm, and quantized using 6 quantization directions.
The database contains 17 different gestures performed by 6 different users. One user performed 20 repetitions of all 17 gesture classes. The remaining 5 users did 10 repetitions of 5 randomly chosen gesture classes. The quantization points were measured for each user individually. The gestures and quantization points are visualized in Fig. \ref{fig:types_gest}.

\subsection{Results}
The proposed gesture recognition system was implemented using Julia version 0.5.0 \cite{DBLP:journals/corr/abs-1209-5145} and a custom-built factor graph toolbox (ForneyLab)\footnote{ForneyLab is a Julia-based Forney-style factor graph toolbox that is under development by the BIASlab team (http://biaslab.org) inside the Electrical Engineering department of Eindhoven University of Technology.}.

As a measure of performance, we use the recognition rate, defined as
\begin{align}
\text{Recognition rate} = \frac{\text{\# correctly classified}}{\text{total \# of samples}}  \,.
\end{align}

We first tested the classifier on the recordings of the user that performed all 17 gesture classes. Five recordings of every gesture class were randomly selected for training, and the remaining 15 recordings are used as a test set. We started by constructing a prior distribution using one recording (randomly selected from the training set) from each gesture class. Following this, the classifier was trained for each class separately and evaluated on the 15 remaining recordings. This experiment was repeated 6 times, each time with a randomly chosen training set and relearned prior distributions.

\begin{figure}[t]
   \centering
  \scalebox{1}{\includegraphics[width=0.485\textwidth]{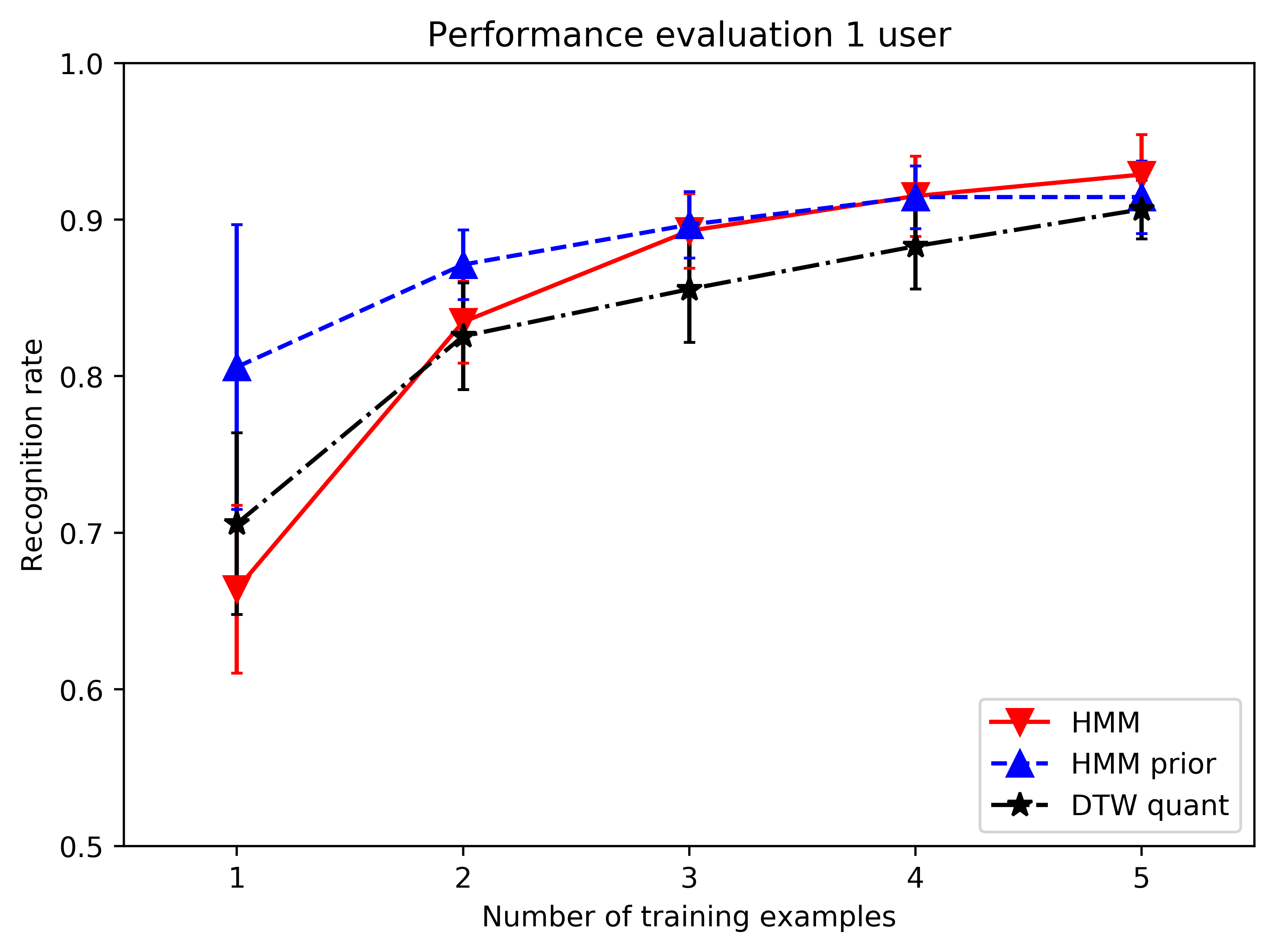}}
\caption{Recognition rates of the proposed algorithm without prior information (HMM), the proposed algorithm with informed prior distributions (HMM prior), the DTW algorithm for one user with 17 gesture classes. The recognition rates are averaged over 6 permutations of the training set. The error bars indicate one standard deviation.}\label{fig:RR_results}
\end{figure}

To test the influence of the constructed prior distribution, we also evaluated the algorithm with an uninformative prior distribution. As a baseline we use the recognition rate of Dynamic Time Warping (DTW). Our version of DTW calculates a minimum distance between two time series, by taking into account that the two time series might be non-linearly warped in time. A gesture is classified to belong to the same class as the training example with minimal distance to the measurement. DTW has been used previously as a classifier for gestures \cite{liu_uwave:_2009, akl_accelerometer-based_2010}. We evaluate DTW on the same quantized signal as the HMMs. Fig. \ref{fig:RR_results} shows the recognition rates of the algorithms.

In view of our interest for one-shot training of the gesture classifier, we are especially interested in classifier performance as a function of number of training examples.
In particular for one training example, the HMM with learned prior distribution clearly outperforms the HMM with uninformative priors and DTW.

For the users that performed 5 different gesture types, the data set is divided into a training set consisting of 3 recordings, and a test set containing the 7 remaining recordings. We started by learning an individual prior distribution for each user by taking one recording from the training set from each gesture class. After this, the classifier was trained and evaluated for each user individually. The experiment was repeated 6 times, with 6 different randomly chosen training sets.

\begin{figure}[t]
   \centering
  \scalebox{1}{\includegraphics[width=0.485\textwidth]{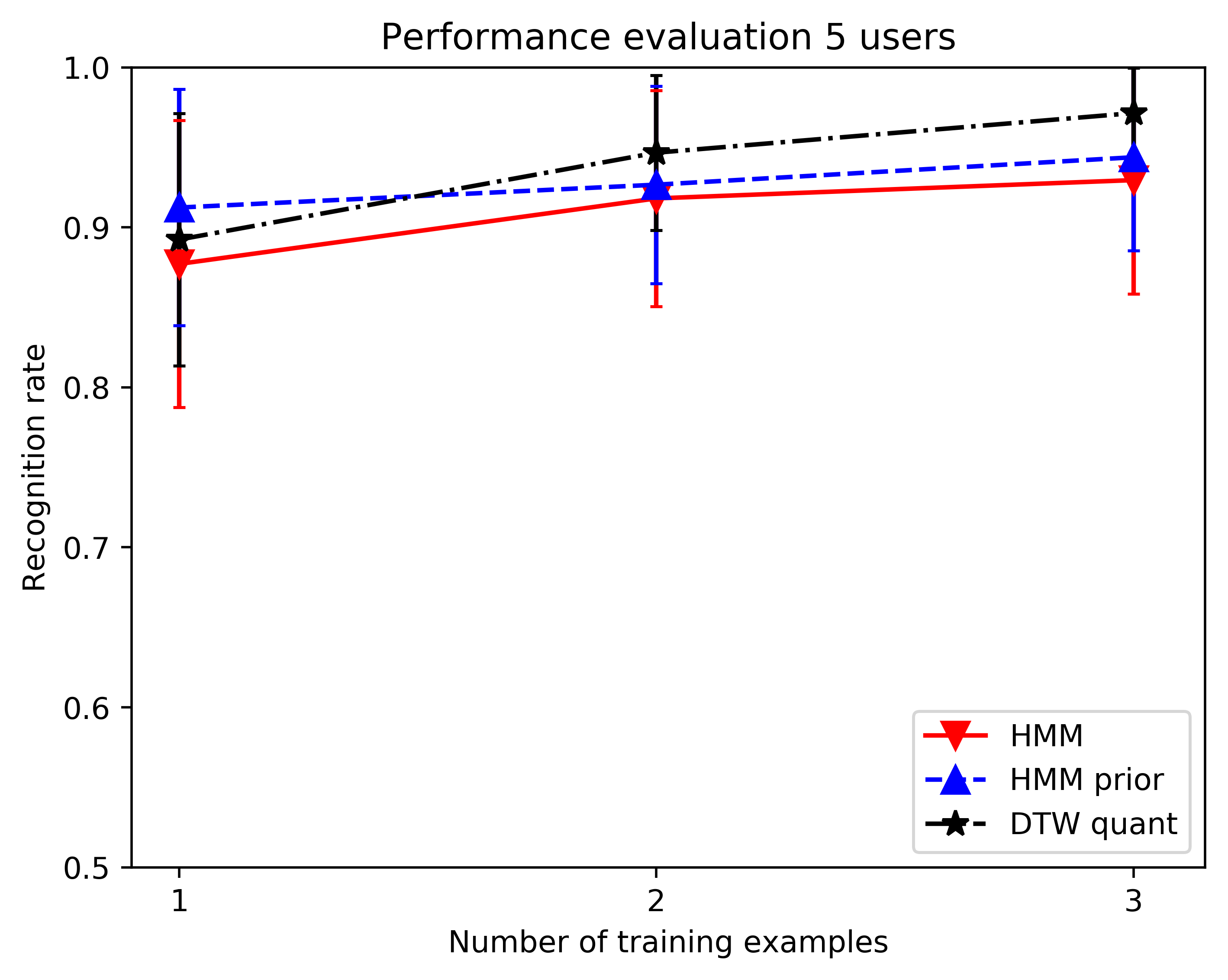}}
\caption{Recognition rates of the DTW algorithm, the proposed algorithm without prior information (HMM), and the proposed algorithm with informed prior distributions (HMM prior) for the 5 users that performed 5 gesture classes. The recognition rate is evaluated per user and for 6 different permutations of the training set. The error bar indicates one standard deviation.}\label{fig:RR_results_others}
\end{figure}

Fig. \ref{fig:RR_results_others} shows the obtained recognition rates together with the results obtained using an uninformative prior and DTW. The experiment was repeated 6 times, with 6 different randomly chosen training sets.

The improvement due to the prior distribution in Fig. \ref{fig:RR_results_others} is smaller than the improvement in Fig.~\ref{fig:RR_results}. It is good to take into account that the number of gesture recordings used to construct the prior distribution is also lower than in Fig. \ref{fig:RR_results}.

There are multiple ways to incorporate these results in a practical gesture recognition system. For example, the prior distribution can be constructed by the developers of the algorithm. Another possibility is to allow users to provide prior distributions themselves. This means that the system will take longer to set up, but when a user wants to learn a specific gesture under in-situ conditions, it will require fewer training examples.

\section{Real-time implementation}\label{sec:on}
The proposed gesture classifier can be extended into a working gesture recognition system that detects and  classifies gestures on ``streaming data'' without end points. We consider two different approaches. In the first approach, the classifier runs on-line (is always ``on'') while updating a  ``localized'' log-likelihood for each trained gesture class. In the second approach a custom algorithm is designed to detect a so-called ``key-gesture" to trigger the proposed classifier.

\subsection{Detection by tracking a localized log-likelihood}\label{ssec:localized-likelihood}
In order to perform time-varying classification in a data stream, the marginal likelihood of Eq.~\ref{eq:marg_approx} cannot be used since it takes all previously seen data points equally into account. Because gestures cover a limited time span, the classifier must update and return a local-in-time likelihood for each class. In \cite{loeliger_localizing_2009}, a localized (time-varying) model log-likelihood is defined for each gesture class by
\begin{equation}
L_t^{(k)} = \gamma L^{(k)}_{t-1} + \log{p(y_t|y_1,...,y_{t-1},\theta_k)}\,,
\end{equation}
where $\gamma \in [0,1]$ is a forgetting factor, $L_{t-1}^{(k)}$ is the localized log-likelihood for the previous time step for class $k$, and $p(y_t|y_1,...,y_{t-1})$ is the likelihood of an observation, which can be evaluated using message passing in the factor graph. In order to make a classification decision, the current likelihood $L_t^{(k)}$ can be tested against a threshold to determine whether a gesture is made.

In our experience, this approach is not a good candidate for implementation of a personalizable gesture recognition system. A user with little knowledge of the operation of the system might easily choose a gesture that is indistinguishable from another common movement. For instance, assume that a user trains a gesture where he moves his arm up and down. In that case, while eating from a bowl of cereal, the measured signals will be very similar to this gesture, due to the up and down movements of the arm. This issue gets exacerbated by the fact that we only measure relative orientations and not exact positions and do not take the surroundings into account, which might have been possible with a vision-based technique.

\subsection{Activation by a key gesture}
\begin{figure}[t]
   \centering
\scalebox{1}{ \includegraphics[width=0.5\textwidth]{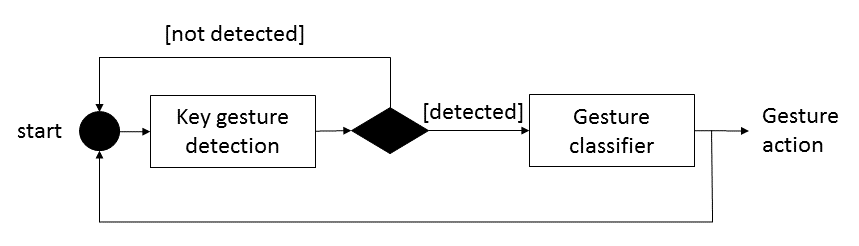}}
\caption{The algorithm for key gesture detection is searching in the data stream for the key gesture. Once this is detected, the proposed gesture classifier is started.}\label{fig:online}
\end{figure}

To overcome the problems related to the first approach, we propose a second method for real-time implementation. In this method the personalized gesture classifier is  triggered by a key gesture (see Fig. \ref{fig:online}), this is done more often in gesture recognition, e.g. in \cite{t._freeman_television_1995}. This key gesture must be computationally cheap to detect and easy to perform, but should be unlikely to trigger by accident.

\begin{figure}[t]
   \centering
 \scalebox{1}{\includegraphics[width=0.485\textwidth]{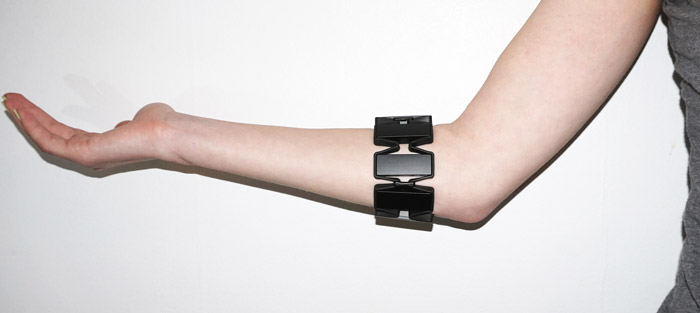}}
\caption{The key gesture that activates the gesture classifier.}\label{fig:key_gesture}
\end{figure}

We have chosen the gesture in Fig. \ref{fig:key_gesture} as our key gesture and designed a custom algorithm to detect this gesture.
From the received orientation data a vector $\mathbf{k}'$ can be extracted that points into the same direction as the palm of the hand. When this vector is pointing upwards for more than one second, the algorithm concludes that the key gesture has been executed and the personalized classifier is started when the arm is turned back.

This proposed detection algorithm for the key gesture is clearly a simple ad-hoc algorithm, but we found it to be cheap and accurate. In general, detection of key gestures could be implemented using more sophisticated methods, e.g., by the localized likelihood method of Sec.~\ref{ssec:localized-likelihood}.

To test the key gesture algorithm, an experiment was conducted in which the user wore the sensor bracelet for one hour while doing desk work. No instances of false positives or false negatives were reported.

\section{Related work}\label{sec:relwork}
Various methods for gesture recognition have been proposed in literature. Many systems are specifically designed for vision-based techniques, e.g. \cite{chen_hand_2003}. Vision-based techniques require a lot of preprocessing and extracting relevant features requires much effort as compared to methods based on inertial and magnetic sensors. Furthermore, it is important that the subject is always illuminated and the background is uniform. We wanted to focus on a portable system that could be used anywhere; therefore, inertial and magnetic sensors were more appropriate for our first tests.
In general, most gesture recognition algorithms are based on dynamic time warping \cite{liu_uwave:_2009, akl_accelerometer-based_2010}, hidden Markov models \cite{mantyla_hand_2000,chen_hand_2003,zhang_framework_2011,kela_accelerometer-based_2006,pylvanainen_accelerometer_2005} or neural networks \cite{wu_deep_2016}. Most of the proposed approaches need many training examples and are therefore not personalizable.

Liu \emph{et al.} \cite{liu_uwave:_2009} also describe a personalizable gesture recognition algorithm for accelerometer data. Their algorithm is based on dynamic time warping and only uses one training example. They have tested their algorithm with  8 different gestures collected from 8 different users and report an accuracy of 93.5 percent using 33 quantization levels.
Cabrera \emph{et al.} \cite{cabrera_embodied_2016} developed a one-shot learning algorithm for gestures collected by a Microsoft Kinect. Instead of designing a specific classifier for one-shot learning, they propose a Gaussian mixture model for gestures from which samples are drawn to train a classifier.
Kela \emph{et al.} \cite{kela_accelerometer-based_2006} also use discrete-output hidden Markov models for accelerometer data. They train their system using the Baum-Welch algorithm for 8 different gestures. Making use of 4 training examples they report an accuracy of over 90 percent. They also mention that for smaller training sets their approach suffers from overfitting and attempt to solve this issue in a later paper by adding Gaussian noise to the training examples \cite{mantyjarvi_enabling_2004}. In the latter paper, a 92 percent recognition rate was obtained for 8 gesture classes, using one training example and one training example with added Gaussian noise.
To the best of our knowledge, no gesture recognition algorithm has  been proposed before that uses variational inference to train ``gesture'' priors in hidden Markov models. Hierarchical Bayesian models for one-shot training have recently been proposed as models for how humans learn language and speech \cite{lake_one-shot_2014}. The idea of sharing prior knowledge in the form of prior distributions is not new, as it is also used (for instance by \cite{fei-fei_bayesian_2003, fei-fei_one-shot_2006} for learning object categories for images.
Regarding our learning methods, an extensive comparison of variational inference and EM for hidden Markov models is provided in \cite{beal_variational_2003}.

\section{Discussion}
\subsection{Gesture classification}
We have proposed a classifier based on a hierarchical probabilistic model that supports one-shot learning and classification of temporal sequences. While the classifier is not constrained to work on any specific type of data, we have tested the classifier on arm gestures. The Bayesian approach facilitates learning of prior knowledge about gestures, which reduces the number of required training examples for new gestures. We exploited this by creating a two-step learning approach. First, a general prior based on a set of measurements without gesture labels is learned. In the second step, the specific gesture classes are learned based on the prior distribution and a few labeled gesture measurements. We tested this approach on a large database, which gave good results compared to competing algorithms in the quantized domain.

Despite the favorable results, the classifier also suffers from some limitations. Due to the limited number of quantization points, the classifier cannot distinguish between gestures that lead to the same quantized measurement sequence. To evaluate the reduction in predictive performance due to quantization, we compared the performances of the DTW classifier with and without measurement quantization. It appears that getting rid of the (course) input quantization stage improves the predictive performance. This suggests that there is room to increase the performance of our classifier as well by optimizing the quantizer.
A straightforward way to do this would be to extend the generative model with a continuous domain observation model. The parameters of this observation model (which imply the quantization grid) could then be learned directly from the continuous measurements, which should result in a reduced quantization loss.

It might also prove to be useful to have a mechanism that prevents users from training multiple new gesture classes that are too similar. A simple approach would be to evaluate the marginal likelihood $p(\mathbf{y}|k)$ of already existing classes on a new gesture measurement $\mathbf{y}$, and warn the user or refuse to generate a new class when this value is (too) high.

\subsection{Measuring gestures}
We decided to test our classifier using directional data measured by the Myo bracelet. Although the classifier might work with any type of data, we still want to note some advantages and disadvantages of using data that is measured with inertial and magnetic sensors.

Especially compared to vision-based approaches (approaches that use cameras to detect movements), inertial and magnetic sensors enjoy a couple of advantages. For instance, these sensor signals do not depend on ambient lighting, and users might perceive it as a lesser violation of privacy. Moreover, these sensors tend to be more mobile, e.g., can be assembled in small wearable devices. On the other hand, measuring gestures using orientation signals also has a number of disadvantages when compared to camera systems. Orientation sensors can only measure rotation and not absolute position (which can be extracted when cameras are used). For this reason, not every gesture can be detected and they need to be restricted in order to ensure that detection is possible. Also, orientation measurements will only be correct when no force is working on the user other than gravity. For example, in an accelerating car, the orientation measurements will not be accurate.

Nevertheless, inertial and magnetic sensors to measure gestures seem to be a very practical solution for a portable system at the moment.

\subsection{Real-time implementation}

We proposed and implemented two methods to detect gestures in a continuous stream of data.

The first method is based on implementing the proposed classifier on-line, using a localized version of the model likelihood. The main problem with this method was that only directional data is not specific enough to detect gestures. A small movement of the arm will lead to the same directional data as a real gesture. This is not a real problem when the system is applied in a setting where arm movements are mostly gestures. However, for the application of gesture recognition that we had in mind, the user uses gesture recognition every now and then to send a command to a device while doing other tasks. The second method that we considered is more suitable for this.

The second method makes use of a predefined ``key gesture" to trigger the classifier. We introduced a key gesture and a method to detect this key gesture. The detection with a key gesture is less sensitive to false positives than the first method. The proposed algorithm works well, however more sophisticated solutions are possible if needed.

Finally, we mention that we have not yet used an important advantage of using generative models, namely the possibility of detecting outliers by evaluating the probability of the data under the learned models. This could be used as a check to determine if the movement following the key gesture is indeed a learned gesture.

\subsection{Further work}
There are several aspects that can be investigated in the future, such as extending the hidden Markov model to process continuously valued data. The generative modeling approach provides a principled framework to make these types of extensions. Moreover, it would be interesting to see the effect of such extensions on one-shot learning.

The recordings used for the construction of the prior distributions were now selected randomly from the training set. This already gives an improved performance as compared to an uninformative prior distribution. However, better performance might be achieved when the recordings used for construction of the prior distribution are selected in a different way, or if the hierarchy of the model is extended further.

The proposed method is quite general and it could be considered to test the method for more applications, e.g., learning and detection of (in-)appropriate walking patterns.

\section{Conclusion}
We have proposed a gesture classifier for directional data of the arm, based on a Bayesian hierarchical model. The Bayesian approach made it possible to include prior information about gestures through prior distributions. We validated the classifier on a database containing 17 different gesture classes, performed by 6 different users. Including prior distributions of gestures was especially advantageous for one-shot learning. We compared our proposed algorithm to a state-of-the-art algorithm for one-shot learning of gestures. Our proposed algorithm had a higher recognition rate than the DTW algorithm on the quantized signal. This might indicate that it might also outperform DTW on the continuous signal, when quantization is removed or the generative model is extended to the continuous domain. Finally, we proposed and implemented the classifier in a real-time fashion using two different methods.

\appendix

\section{Notation}\label{app:notation}
Our notation is based on the notation of \cite{bishop_pattern_2006}. Vectors are indicated by bold lowercase letters, e.g. $\mathbf{y}$. Matrices are denoted by by bold uppercase letters, like $\mathbf{A}$. A row vector with T elements is written as $(x_1,...,x_T)$. With the notation $\{1,2,...,N\}$ a set is meant which would be represented in set-builder representation as: $\{x\in \mathbb{N}| x\leq N\}$. For quaternions the notation of \cite{horn_closed-form_1987} is used, in which quaternions are denoted using bold lowercase letters with a dot, $\mathbf{\dot{q}}$.

\section{Variational message passing for learning}
\label{sec:var}
In this section, we specify the variational message passing rules for inference of the parameter distributions.
For computational simplicity during estimation of the parameter distributions, one node is made for the complete ``dynamical model" in Eq.~ \ref{eq:compgenmod}, we call this node the ``HM" node. There is no edge for $\mathbf{x}$; the needed quantities are calculated internally.
\begin{figure}[htb]
   \centering
    \scalebox{1}{

\begin{tikzpicture}
        [node distance=15mm,auto,>=stealth']
        \begin{scope}
           \node[box, node distance=15mm] (ex_node){$HM$};
	\node[smallbox,above of = ex_node, node distance = 15 mm](pi){};
	\node[clamped,below of = ex_node, node distance = 15 mm](X){};
	\node[smallbox,right of = ex_node, node distance = 15 mm](A){};
	\node[smallbox, left of = ex_node, node distance = 15 mm](C){};
	
	\node[below of = C, node distance = 5mm](C_l){$p(\mathbf{C})$};
	\node[below of = A, node distance = 5mm](A_l){$p(\mathbf{A})$};
	\node[right of = pi, node distance = 7mm](pi_l){$p(\boldsymbol{\pi})$};
         
	\path[line] (C) edge[-]node[anchor=south]{$\mathbf{C}$}  (ex_node); 
	\path[line] (X) edge[-]node[anchor=west]{$\mathbf{y}^{(k)}=(y_1^{(k)},...,y_T^{(k)})$} (ex_node); 
	\path[line] (A) edge[-]node[anchor=south]{$\mathbf{A}$} (ex_node);           
	\path[line] (pi) edge[-]node[anchor=west]{$\boldsymbol{\pi}$} (ex_node);   
        \end{scope}
    \end{tikzpicture}
}
\caption{Factor graph used for estimation of the parameter distributions.}\label{fig:learning_node}
\end{figure}
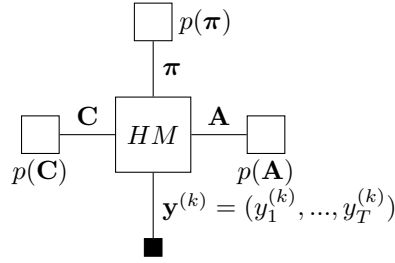

The variational parameter distributions we want to estimate during learning, $q(\boldsymbol{\pi})$, $q(\mathbf{A})$ and $q(\mathbf{C})$, can each be split into a product of a forward and a backward message:
\begin{align*}
q(\boldsymbol{\pi}) &= \overrightarrow{\nu}(\boldsymbol{\pi})\overleftarrow{\nu}(\boldsymbol{\pi})\\
q(\mathbf{A}) &= \overrightarrow{\nu}(\mathbf{A})\overleftarrow{\nu}(\mathbf{A}) \\
q(\mathbf{C}) &= \overrightarrow{\nu}(\mathbf{C})\overleftarrow{\nu}(\mathbf{C})\,.
\end{align*}
The messages $\overleftarrow{\nu}(\boldsymbol{\pi})$, $\overleftarrow{\nu}(\mathbf{A})$ and $\overleftarrow{\nu}(\mathbf{C})$ depend on the incoming messages from the rest of the factor graph. For the case of Fig. \ref{fig:learning_node} they are equal to $p(\boldsymbol\pi)$, $p(\mathbf{A})$ and $p(\mathbf{C})$. The messages $\overrightarrow{\nu}(\boldsymbol\pi)$, $\overrightarrow{\nu}(\mathbf{A})$, $\overrightarrow{\nu}(\mathbf{C})$ are determined by the update rules of the node itself. The update rules are based on the algorithm of MacKay \cite{mackay_ensemble_1997} and are modified slightly to fit in the factor graph framework.

The update messages for the node are:
\begin{align*}
\overrightarrow{\nu}(\boldsymbol{\pi}) &= \operatorname{Dir}{(\boldsymbol{\pi}|\mathbf{w}^{(\boldsymbol{\pi})})}\,,
\end{align*}
where the elements of $\mathbf{w}^{(\boldsymbol{\pi})} \in \mathbb{R}^M$ are defined as $w^{(\boldsymbol{\pi})}_{i} = \sum_{\mathbf{x}} q(\mathbf{x}) \delta(x_1=i)$.
\begin{align*}
\overrightarrow{\nu}(\mathbf{A}) &= \prod^M_{j=1} \operatorname{Dir}((A_{1j},...,A_{Mj})|(W_{1j}^{(\mathbf{A})},...,W_{Mj}^{(\mathbf{A})}))\,,
\end{align*}
where $W^{(\mathbf{A})}_{ij}=\sum_{\mathbf{x},t} q(\mathbf{x})\delta(x_{t+1} = i, x_{t} = j)$.
\begin{align*}
\overrightarrow{\nu}(\mathbf{C}) &= \prod^M_{j=1} \operatorname{Dir}{((C_{1j},...,C_{Nj})|( W^{(\mathbf{C})}_{1j},...,W^{(\mathbf{C})}_{Nj}))}\,,
\end{align*}
where $W^{(\mathbf{C})}_{ij}=\sum_{\mathbf{x},t} q(\mathbf{x})\delta(y_t = i, x_t = j)$.

Since $q(\mathbf{x})$ is not implemented using message passing, it stays the same as in \cite{mackay_ensemble_1997}:
\begin{align*}
q(\mathbf{x}) \propto \pi^*_{x_1}\prod^{T-1}_{t=1}A^{*}_{x_{t+1}x_t}\prod^T_{t=1} C^{*}_{y_tx_t}\,,
\end{align*}
where $A^*_{x_{t+1}x_t}$, $C^*_{y_tx_t}$, and $\pi^*_{x_1}$ are defined as:
\begin{align*}
A^*_{x_{t+1}x_t} &= \exp{\int q(\mathbf{A})\log{A_{x_{t+1}x_t}}\mathbf{\d A}}\\
C^*_{y_tx_t} &= \exp{\int q(\mathbf{C}) \log{C_{y_tx_t}}\mathbf{\d C}}\\
\pi^*_{x_1} &= \exp{\int q(\boldsymbol{\pi})\log{\pi_{x_1}}\operatorname{d}\!\boldsymbol{\pi}}\,.
\end{align*}
Note that the update rule of $q(\mathbf{x})$ depends on $q(\mathbf{A})$, $q(\mathbf{C})$, and $q(\boldsymbol{\pi})$; and the update rules of $\overrightarrow{\nu}(\mathbf{A})$, $\overrightarrow{\nu}(\mathbf{C})$, and $\overrightarrow{\nu}(\mathbf{C})$ depend on $q(\mathbf{x})$ only. The relevant properties of $q(\mathbf{x})$ can therefore be evaluated using the forward-backward algorithm, which thereafter can be used to update $\overrightarrow{\nu}(\mathbf{A})$, $\overrightarrow{\nu}(\mathbf{C})$, and $\overrightarrow{\nu}(\boldsymbol\pi)$. This has to be repeated until convergence.

\section*{Acknowledgments}
The authors would like to thank the BIASlab team members (\url{https://biaslab.github.io/member/}) for their help during this project and for participating in the experiment.

\clearpage
\bibliographystyle{abbrv}
\bibliography{ieeetr}

\begin{thebibliography}{10}

\bibitem{database}
A.~van~Diepen, M.~G.~H.~Cox and A.~de~Vries.
\newblock Orientation of the arm used for Gesture Recognition.
\newblock In {4TU.Centre for Research Data}, 2020.

\bibitem{noauthor_myo_nodate}
Myo {Gesture} {Control} {Armband}.

\bibitem{akl_accelerometer-based_2010}
A.~Akl and S.~Valaee.
\newblock Accelerometer-based gesture recognition via dynamic-time warping,
  affinity propagation, \& compressive sensing.
\newblock In {\em 2010 {IEEE} {International} {Conference} on {Acoustics},
  {Speech} and {Signal} {Processing}}, pages 2270--2273, Mar. 2010.

\bibitem{beal_variational_2003}
M.~J. Beal.
\newblock Variational algorithms for approximate {Bayesian} inference.
\newblock {\em Ph.D. Thesis}, 2003.

\bibitem{DBLP:journals/corr/abs-1209-5145}
J.~Bezanson, S.~Karpinski, V.~B. Shah, and A.~Edelman.
\newblock Julia: {A} fast dynamic language for technical computing.
\newblock {\em CoRR}, abs/1209.5145, 2012.

\bibitem{bishop_pattern_2006}
C.~M. Bishop.
\newblock {\em Pattern recognition and machine learning}.
\newblock Information science and statistics. Springer, New York, 2006.

\bibitem{cabrera_embodied_2016}
M.~E. Cabrera and J.~P. Wachs.
\newblock Embodied gesture learning from one-shot.
\newblock In {\em 2016 25th {IEEE} {International} {Symposium} on {Robot} and
  {Human} {Interactive} {Communication} ({RO}-{MAN})}, pages 1092--1097, Aug.
  2016.

\bibitem{campbell_invariant_1996}
L.~W. Campbell, D.~A. Becker, A.~Azarbayejani, A.~F. Bobick, and A.~Pentland.
\newblock Invariant features for 3-{D} gesture recognition.
\newblock In {\em , {Proceedings} of the {Second} {International} {Conference}
  on {Automatic} {Face} and {Gesture} {Recognition}, 1996}, pages 157--162,
  Oct. 1996.

\bibitem{chen_hand_2003}
F.-S. Chen, C.-M. Fu, and C.-L. Huang.
\newblock Hand gesture recognition using a real-time tracking method and hidden
  {Markov} models.
\newblock {\em Image and Vision Computing}, 21(8):745--758, Aug. 2003.

\bibitem{dauwels_variational_2007}
J.~Dauwels.
\newblock On {Variational} {Message} {Passing} on {Factor} {Graphs}.
\newblock In {\em 2007 {IEEE} {International} {Symposium} on {Information}
  {Theory}}, pages 2546--2550, June 2007.

\bibitem{fei-fei_bayesian_2003}
L.~Fei-Fei, R.~Fergus, and P.~Perona.
\newblock A {Bayesian} {Approach} to {Unsupervised} {One}-{Shot} {Learning} of
  {Object} {Categories}.
\newblock In {\em Proceedings of the {Ninth} {IEEE} {International}
  {Conference} on {Computer} {Vision} - {Volume} 2}, {ICCV} '03, pages 1134--,
  Washington, DC, USA, 2003. IEEE Computer Society.

\bibitem{fei-fei_one-shot_2006}
L.~Fei-Fei, R.~Fergus, and P.~Perona.
\newblock One-shot learning of object categories.
\newblock {\em IEEE Transactions on Pattern Analysis and Machine Intelligence},
  28(4):594--611, Apr. 2006.

\bibitem{forney_codes_2001}
G.~D. Forney.
\newblock Codes on graphs: normal realizations.
\newblock {\em IEEE Transactions on Information Theory}, 47(2):520--548, Feb.
  2001.

\bibitem{horn_closed-form_1987}
B.~K.~P. Horn.
\newblock Closed-form solution of absolute orientation using unit quaternions.
\newblock {\em JOSA A}, 4(4):629--642, Apr. 1987.

\bibitem{horsley_wave_2016}
D.~Horsley.
\newblock Wave hello to the next interface.
\newblock {\em IEEE Spectrum}, 53(12):46--51, Dec. 2016.

\bibitem{jordan_introduction_1999}
M.~I. Jordan, Z.~Ghahramani, T.~S. Jaakkola, and L.~K. Saul.
\newblock An {Introduction} to {Variational} {Methods} for {Graphical}
  {Models}.
\newblock {\em Mach. Learn.}, 37(2):183--233, Nov. 1999.

\bibitem{kela_accelerometer-based_2006}
J.~Kela, P.~Korpip{\"a}{\"a}, J.~M{\"a}ntyj{\"a}rvi, S.~Kallio, G.~Savino,
  L.~Jozzo, and S.~D. Marca.
\newblock Accelerometer-based gesture control for a design environment.
\newblock {\em Personal and Ubiquitous Computing}, 10(5):285--299, Aug. 2006.

\bibitem{lake_one-shot_2014}
B.~M. Lake, C.-y. Lee, J.~Glass, and J.~B. Tenenbaum.
\newblock One-shot learning of generative speech concepts.
\newblock 2014.

\bibitem{liu_uwave:_2009}
J.~Liu, L.~Zhong, J.~Wickramasuriya, and V.~Vasudevan.
\newblock {uWave}: {Accelerometer}-based personalized gesture recognition and
  its applications.
\newblock {\em Pervasive and Mobile Computing}, 5(6):657--675, Dec. 2009.

\bibitem{loeliger_introduction_2004}
H.~A. Loeliger.
\newblock An introduction to factor graphs.
\newblock {\em IEEE Signal Processing Magazine}, 21(1):28--41, Jan. 2004.

\bibitem{loeliger_localizing_2009}
H.~A. Loeliger, L.~Bolliger, C.~Reller, and S.~Korl.
\newblock Localizing, forgetting, and likelihood filtering in state-space
  models.
\newblock In {\em 2009 {Information} {Theory} and {Applications} {Workshop}},
  pages 184--186, Feb. 2009.

\bibitem{loeliger_factor_2007}
H.~A. Loeliger, J.~Dauwels, J.~Hu, S.~Korl, L.~Ping, and F.~R. Kschischang.
\newblock The {Factor} {Graph} {Approach} to {Model}-{Based} {Signal}
  {Processing}.
\newblock {\em Proceedings of the IEEE}, 95(6):1295--1322, June 2007.

\bibitem{mackay_ensemble_1997}
D.~J.~C. MacKay.
\newblock Ensemble {Learning} for {Hidden} {Markov} {Models}.
\newblock Technical report, 1997.

\bibitem{mantyjarvi_enabling_2004}
J.~M{\"a}ntyj{\"a}rvi, J.~Kela, P.~Korpip{\"a}{\"a}, and S.~Kallio.
\newblock Enabling {Fast} and {Effortless} {Customisation} in {Accelerometer}
  {Based} {Gesture} {Interaction}.
\newblock In {\em Proceedings of the 3rd {International} {Conference} on
  {Mobile} and {Ubiquitous} {Multimedia}}, {MUM} '04, pages 25--31, New York,
  NY, USA, 2004. ACM.

\bibitem{mantyla_hand_2000}
V.~M. Mantyla, J.~Mantyjarvi, T.~Seppanen, and E.~Tuulari.
\newblock Hand gesture recognition of a mobile device user.
\newblock In {\em 2000 {IEEE} {International} {Conference} on {Multimedia} and
  {Expo}. {ICME}2000. {Proceedings}. {Latest} {Advances} in the {Fast}
  {Changing} {World} of {Multimedia} ({Cat}. {No}.00TH8532)}, volume~1, pages
  281--284 vol.1, 2000.

\bibitem{pylvanainen_accelerometer_2005}
T.~Pylv{\"a}n{\"a}inen.
\newblock Accelerometer {Based} {Gesture} {Recognition} {Using} {Continuous}
  {HMMs}.
\newblock In {\em Pattern {Recognition} and {Image} {Analysis}}, pages
  639--646. Springer, Berlin, Heidelberg, June 2005.

\bibitem{rabiner_tutorial_1989}
L.~Rabiner.
\newblock A {Tutorial} on {Hidden} {Markov} {Models} and {Selected}
  {Applications} on {Speech} {Recognition}.
\newblock {\em Proceedings of the IEEE}, 77:257--286, Mar. 1989.

\bibitem{t._freeman_television_1995}
W.~T.~Freeman and C.~D.~Weissman.
\newblock Television {Control} by {Hand} {Gestures}.
\newblock In {\em Proceedings of the {International} {Workshop} on {Automatic}
  {Face}- and {Gesture}- {Recognition}.}, Zurich, June 1995.

\bibitem{wu_deep_2016}
D.~Wu, L.~Pigou, P.~J. Kindermans, N.~D.~H. Le, L.~Shao, J.~Dambre, and J.~M.
  Odobez.
\newblock Deep {Dynamic} {Neural} {Networks} for {Multimodal} {Gesture}
  {Segmentation} and {Recognition}.
\newblock {\em IEEE Transactions on Pattern Analysis and Machine Intelligence},
  38(8):1583--1597, Aug. 2016.

\bibitem{yamato_recognizing_1992}
J.~Yamato, J.~Ohya, and K.~Ishii.
\newblock Recognizing human action in time-sequential images using hidden
  {Markov} model.
\newblock In {\em Proceedings 1992 {IEEE} {Computer} {Society} {Conference} on
  {Computer} {Vision} and {Pattern} {Recognition}}, pages 379--385, June 1992.

\bibitem{zhang_framework_2011}
X.~Zhang, X.~Chen, Y.~Li, V.~Lantz, K.~Wang, and J.~Yang.
\newblock A {Framework} for {Hand} {Gesture} {Recognition} {Based} on
  {Accelerometer} and {EMG} {Sensors}.
\newblock {\em IEEE Transactions on Systems, Man, and Cybernetics - Part A:
  Systems and Humans}, 41(6):1064--1076, Nov. 2011.

\end{thebibliography}

\end{document}